\documentclass[pra,preprint,tightenlines,showpacs,11pt]{revtex4} 
\usepackage{epsfig} 

\begin{document}  
    
\title{ On the dynamics of transfer-ionization in fast ion-atomic collisions  }
 
\author{ A.B.Voitkiv }   
\affiliation{ Max-Planck-Institut f\"ur Kernphysik, 
Saupfercheckweg 1, D-69117 Heidelberg, Germany } 
\author{ X.Ma }  
\affiliation{ Institute of Modern Physics, Chinese Academy of Science, 
730000 Lanzhou, China }


\begin{abstract} 

We consider trasfer-ionization in collisions of fast 
($3.6$ -- $11$ MeV/u) protons, alpha-particles and lithium nuclei 
with helium atoms. There are just a few basic mechanisms 
contributing to this process which can be grouped into correlated ones, 
which crucially depend on the electron-electron interaction, 
and uncorrelated, which do not need this interaction to proceed. 
We show that by exploring momentum spectra of the emitted electrons 
the correlated and uncorrelated mechanisms can be cleary separated 
from each other. This exploration also enables one 
to get insight into subtle details of the dynamics of transfer-ionization.  
 
\end{abstract} 


\maketitle 



\section{Introduction} 

Ionization and electron transfer (electron capture), 
which may occur in collisions between an atom and a bare nucleus, 
represent two of basic collision processes studied by atomic physics. 
In the process of ionization the atom emits an electron, which 
after the colllision moves freely in space, while 
in the transfer process an electron 
initially bound in the atom is captured   
into a bound state of the moving ion. 
Both of these processes possess interesting physics, 
their study is of importance for many applications 
and various aspects of these processes 
have been attracting attention for decades. 

Quite an interesting situation is encountered 
when a combination of transfer and ionization occurs 
in a single collision event. Such a process,  
which becomes possible if the atomic target 
has at least two electrons, is called transfer-ionization. 
During the last decade transfer-ionization 
in collisions of protons with helium atoms  
has attracted much attention \cite{mergel}-\cite{schmidt}. 

There are only a few known basic mechanisms 
governing transfer-ionization in fast colllisions. 
Depending on whether the electron-electron 
interaction (correlations) plays 
in them a crucial role or not, these mechanisms 
can be subdivided into "correlated" 
and "uncorrelated" ones. 

The group of uncorrelated mechanisms consists of 
the so called independent transfer-ionization 
(ITI) and capture--shake-off (C-SO). 
In the ITI electron capture and emission occur 
due to the "independent" interactions 
of the electrons with the ionic projectile.    
In a theoretical description this mechanism appears 
starting with second order perturbation theory 
in the ion-atom interaction  
and for its realization does not need  
any electron-electron interaction.  
 
According to the C-SO mechanism,  
a fast removal of the captured electron from 
the atom leads to a "sudden" change  
of the atomic potential in which 
the other electron is moving. 
As a result, the electron tries to adjust 
its state to the new potential and 
has a nonzero probability 
to become unbound \cite{mcg}.  
 
The correlated mechanisms are more interesting. 
They include so called electron-electron Thomas (EET) mechanism 
and a mechanism which will be termed here as electron-electron Auger (EEA). 
According to the EET, transfer-ionization proceeds 
in two steps \cite{thomas}, \cite{briggs}. 
In the first step, as a result of a binary collision with 
the ion, one of the electrons acquires 
a velocity $\sqrt{2} v$, where $v$ is the ion velocity, 
moving under the angle of $45^0$ with respect 
to the motion of the ion. 
In the second step this electron scatters on 
the other electron acquiring, as simple kinematics shows, 
a velocity equal to the projectile velocity,   
both in absolute magnitude and direction,  
that makes the capture very probable. 
The same kinematics also tells that the other electron 
in this process gets a velocity, which is perpendicular 
to the projectile velocity and whose absolute value is equal to $v$. 
Thus, as a result of the EET one electron is captured and 
the other is emitted perpendicular to the projectile motion. 

The electron-electron interaction is also 
the (main) driving force of the EEA mechanism. 
The physics of the latter becomes very transparent 
when it is viewed in the rest frame of the projectile nucleus. 
The functioning of this mechanism is based 
on the fact that the presence of the 
second nucleus makes the initial configuration of atomic particles 
unstable with respect to a kind of Auger decay. 
Indeed, in the presence of this nucleus 
one of the electrons, which initially belongs 
to a bound configuration of fast moving particles constituting 
the atom, undergoes a radiationless transition 
\cite{foot_note} into a bound state of the ion by transferring 
(most of) energy excess to the another atomic electron which, 
as a result of this, is emitted from the atom 
\cite{we-EE}, \cite{ich-EE}. 
A clear signature of this mechanism is that 
in the rest frame of the atom the electron is emitted 
in the direction opposite to 
the projectile motion 
\cite{we-EE}, \cite{ich-EE}, \cite{daniel}. 

One has to emphasize that the mechanisms for 
transfer-ionization, discussed above, are 
in essence high-energy approximations,  
the validity of which improves with increasing 
impact energy. Therefore, the description 
of transfer-ionization in terms of 
these mechanisms becomes really meaningful 
only provided the collision velocity is high enough:  
$v \gg v_i, v_f $, where $v_i \sim Z_t$ and $v_f \sim Z_p$  
the typical velocities of the electron(s) in the initial 
and final bound states, respectively, and $Z_t$ ($Z_p$) 
is the charge of the nucleus of the target (projectile). 
This implies that in order to get an insight into the physics 
of transfer-ionization by considering this process as driven 
by these mechanisms, even in collisions with protons 
the impact velocity should lie in the range  
$ v \stackrel{>}{\sim} 10 v_0$, where $ v_0$ is 
the Bohr velocity in atomic hydrogen. 

Although transfer-ionization was studied in a number of papers, 
most of them were concerned with the total cross section. 
A better understanidng of the physics of this process 
can be obtained when differential cross sections are
explored. Concerning such cross sections in the case 
of transfer-ionization in fast collisions  
only the cross sections singly differential 
in the momentum component of the emitted electron 
or the target recoil ion, parallel/antiparallel 
to the projectile velocity, have been considered  
(see e.g. \cite{schmidt}, \cite{we-EE}-\cite{ich-EE}).   

However, the exploration of such singly differential 
cross sections even in principle can hardly allow one 
to clearly separate the contributions of 
the correlated and uncorrelated mechanisms 
(and thus to study and understand them better). 
Compared to the singly differential cross sections 
the doubly differential cross sections, which are a function of 
both parallel and perpendicular to the projectile velocity 
components of the momentum of the emitted electron, 
can yield much more information about the process. 
Therefore, in the present paper we consider 
such cross sections for transfer-ionization in collisions of 
fast protons, alpha-particles and lithium nuclei with helium atoms.  
It will, in particular, be shown 
that the study of such doubly differential cross sections 
may enable one to clearly separate and 
identify the different mechanisms contributing 
to transfer-ionization and to get 
a better insight into the physics of this process.  

One should say that all the previous experimental 
studies devoted to the spectra of electrons emitted in transfer-ionization 
were dealing with relatively low impact velocities where, 
as was already mentioned, the discussion of this process 
in terms of the four mechanisms may not yet be very meaningful. 
Therefore. we hope that the present article could  
trigger the interest of experimental physicists 
to the exploration of this process at higher impact velocities.  

Atomic units ($\hbar = m_e = |e| =1 $) are used throughout the paper 
except where the otherwise stated.   

\section{General Consideration}  

In our description of transfer-ionization  
the correlated and uncorrelated mechanisms 
shall be treated separately (and added in the cross section 
incoherently). We begin with 
considering the correlated ones. 

\subsection{The EEA and EET mechanisms} 

The (approximate) transition amplitude for transfer-ionization 
can be written  
\begin{eqnarray} 
a_{fi} = -i \int_{-\infty}^{+\infty} dt %
\langle \Psi_f(t) |\hat{W} |\Psi_i(t)\rangle.      
\label{e1} 
\end{eqnarray}  
Here $\hat{W}$ is the coulomb interaction between 
the electrons and $\Psi_i(t)$ and $\Psi_f(t)$ 
are the initial and final states of the electrons. 

In the nonrelativistic domain of atomic collisions 
the description of electron capture is covariant 
under a Galilean transformation and 
any Galilean frame can be chosen to consider this process. 
Assuming that the target atom is (initially) 
at rest in the laboratory frame,  
we take for the moment the rest frame of the atom    
as our reference frame and choose its origin  
at the position of the atomic nucleus.  
We denote the coordinates of the electrons 
by ${\bf r}_1$ and ${\bf r}_2$. 
The projectile-nucleus with a charge $Z_p$ 
is assumed to move along 
a straight-line trajectory 
${\bf R}(t)= {\bf b} + {\bf v} t$, where 
${\bf b}$ is the impact parameter,  
${\bf v}$ the collision velocity and $t$ the time.   
The coordinates of the 'first' and 'second' 
electrons with respect to the position 
of the projectile are denoted by 
${\bf s}_1$ and ${\bf s}_2$, respectively 
(${\bf s}_j={\bf r}_j-{\bf R}(t)$; $j=1,2$).  

We choose the initial state as   
\begin{eqnarray}
\Psi_i(t) &=&  \Lambda_i \, \varphi_i({\bf r}_1,{\bf r}_2) %
\exp(-i E_i t ).     
\label{e2}  
\end{eqnarray}  
In Eq.(\ref{e2}) $\varphi_i$ is the initial 
unperturbed two-electron atomic state with 
an energy $E_i$ and $ \Lambda_i$ is 
a factor which accounts for the distortion 
of the initial atomic state by the field of 
the incident ion, its form shall be specified later. 
We approximate the state $\varphi_i$ according to   
\begin{eqnarray}
\varphi_i({\bf r}_1,{\bf r}_2) = %
A_i \left( \exp\left(-\alpha r_1 - \beta r_2 \right) %
+ \exp\left(-\alpha r_2 - \beta r_1\right) \right) %
\exp\left( \gamma r_{12}\right),       
\label{e3}  
\end{eqnarray}  
where $A_i$ is the normalization factor, 
$r_{12}=\mid {\bf r}_1 - {\bf r}_2 \mid $ 
is the inter-electron distance and 
the parameters $\alpha$, $\beta$ and $\gamma$ 
can be chosen from the following sets:   
(i) $\alpha=\beta=2 $, $\gamma=0$; 
(ii) $\alpha=\beta=1.69$, $\gamma=0$; 
(iii) $\alpha=\beta=1.86$, $\gamma=0.254$; 
(iv) $\alpha=2.18$, $\beta=1.18$, $\gamma=0$; 
and (v) $\alpha=2.21$, $\beta=1.44$, $\gamma=0.207$.  

The final state is taken according to 
\begin{eqnarray} 
\Psi_f(t) &=& \Lambda_f \frac{1}{\sqrt{2}} \left[  %
\chi_f({\bf s}_1) \exp( i {\bf v} \cdot {\bf r}_1 ) %
\, \phi_{\bf k}({\bf r}_2) %
+ \chi_f({\bf s}_2) 
\exp( i {\bf v} \cdot {\bf r}_2 ) 
\, \phi_{\bf k}({\bf r}_1) \right] %
\nonumber \\   
&& \times \exp(-i (\epsilon_k + \varepsilon_f) t)  %
\exp\left( -i \frac{v^2}{2} t \right).     
\label{e4}    
\end{eqnarray}  
Here, $\chi_f$ is the  
final (unperturbed) bound state of the electron 
captured by the projectile,  
$\varepsilon_f$ the energy of this state 
(as viewed in the rest frame of the projectile) 
and $ \exp\left( i {\bf v} \cdot {\bf r}_j -i v^2 t/2 \right)$ 
the so called translational factor. 
Further, $\phi_{\bf k}$ denotes 
the state of the emitted electron
which moves in the field of the residual 
atomic ion with (asymptotic) momentum ${\bf k}$ and 
energy $\epsilon_k = k^2/2$ and $\Lambda_f$ describes 
the distortions of the states of captured and emitted 
electrons by the fields of the residual atomic ion 
and projectile, respectively. 				

Now we turn to the discussion of the form of 
the distortion factors $\Lambda_i$ and $\Lambda_f$. 
Let us remind the reader that in this paper 
we consider only collisions at high impact velocities 
in which one has $Z_p/v \ll 1$. 
Besides, as will be seen below, in the transfer-ionization process 
the emitted electron has a high velocity ($\sim v \gg Z_p$) 
with respect to the projectile. 
From the work on atomic ionization 
it is known that in such collisions 
the account of the distortion 
does not noticeably changes the result. 
At the same time it is also known that 
for electron transfer reactions 
the effect of the distortion in general 
remains very important even at $Z_p/v \ll 1$. 
Therefore, in our treatment we shall ignore the distortions  
for that electron, which is to be emitted, 
and account only for the distortions acting 
on that electron which is to be captured. 

With such an assumption one can show that 
the transition amplitude in the momentum space, 
\begin{eqnarray} 
S_{fi}({\bf q}_{\perp}) = \frac{1}{2 \pi} %
\int d^2 {\bf b} a_{fi}({\bf b}) 
\exp(i {\bf q}_{\perp} \cdot {\bf b}),       
\label{e5} 
\end{eqnarray}  
is given by  
\begin{eqnarray} 
S_{fi}({\bf q}_{\perp}) = 
S^{\alpha, \beta}_{fi}({\bf q}_{\perp}) 
+ S^{\beta, \alpha}_{fi}({\bf q}_{\perp}).        
\label{e6} 
\end{eqnarray}  
Here, 
\begin{eqnarray} 
S^{\alpha, \beta}_{fi}({\bf q}_{\perp}) &=& 
- \frac{\sqrt{2} i A_i}{ (2 \pi)^3 v}  
\int d^3 {\bf s} \chi^*_f({\bf s}) 
\exp(i {\bf q} \cdot {\bf s}) \Lambda_i({\bf s}) 
\int d^3 \mbox{\boldmath$\kappa$}  
\frac{G_{ion}(\mbox{\boldmath$\kappa$}; \beta)}{\kappa^2 + \gamma^2} 
\nonumber \\ 
&& \times \int d^3 {\bf r} \Lambda^*_f({\bf r}) 
\exp(-i ({\bf v} + {\bf q} + \mbox{\boldmath$\kappa$}) 
\cdot {\bf r}) \exp(-\alpha r)
\label{e7} 
\end{eqnarray}  
where  
\begin{eqnarray} 
G_{ion}(\mbox{\boldmath$\kappa$}; \beta) = 
\int d^3 {\bf r} \phi^*_{\bf k}({\bf r}) 
\exp( i \mbox{\boldmath$\kappa$} \cdot {\bf r})
\exp(-\beta r) 
\label{e8} 
\end{eqnarray}  
and  
\begin{eqnarray} 
{\bf q} = \left({\bf q}_{\perp}, 
\frac{ E_i - \varepsilon_f - k^2/2 - v^2/2 }{v} \right)  
\label{e9} 
\end{eqnarray}  
is the momentum transfer in the collision. 
Note that $S^{\beta, \alpha}_{fi}({\bf q}_{\perp})$ is obtained 
from $S^{\alpha, \beta}_{fi}({\bf q}_{\perp})$ 
by interchanging $\alpha $ and $ \beta $ 
in Eqs. (\ref{e7})-(\ref{e8}). 

The explicit form of the distortion factors 
is taken according to the continuum-distorted-state (CDW) 
model which has been proved quite successful in describing 
the total cross section for capture in a three-body collision 
system (one active electron + two nuclei). 
In this model the distortion factors read  
\begin{eqnarray}
\Lambda_i({\bf s}) &=&  
N(\nu_p) \, \left._1F_1 \right. %
\left( i\nu_p,1,i v s + i {\bf v} \cdot {\bf s} \right) %
\nonumber \\ 
\Lambda_f({\bf r}) &=& N^*(\nu_t) \, %
\left._1F_1 \right. %
\left( -i\nu_t,1, - i v r - i {\bf v} \cdot {\bf r} \right),  
\label{e10} 
\end{eqnarray} 
where $ N(\nu)= e^{ \pi \nu /2 } \Gamma(1-i\nu)$, 
$\nu_p=Z_p/v$, $\nu_t=Z_t/v$, 
and $\Gamma$ and $ \left._1F_1 \right.$  
are the gamma and confluent 
hypergeometric functions, 
respectively (see e.g. \cite{Ab-St}). 

The inclusion of the distortion factor for 
the initial state in the form given by 
the first line of (\ref{e10}) means that in our treatment 
the electron, which is to be transferred, 
in its initial state moves not only in 
the field of the atom 
but also in the (coulomb) field of the projectile.  
Therefore, with such a factor 
the transition amplitude (\ref{e1}) 
describes both the EEA and EET capture channels 
while when this factor is set to unity,   
$\Lambda_i = 1$, the calculated contribution 
of the EEA mechanism becomes much larger but 
the EET mechanism simply "vanishes". 

To conclude this subsection let us note that 
the account of the distortion for the final state turned out 
to be not so crucial. 
Indeed, in cases tested the difference 
between results obtained with the distortion factor $\Lambda_f$
in the form given by the second line of (\ref{e10}) and 
by setting $\Lambda_f = 1$ was not substantial. 
Therefore, taking into account that the neglect 
of this distortion greatly simplifies the calculation 
reducing the computation time, 
in what follows we shall report only results obtained 
when we suppose that $\Lambda_f =1 $. 
 
\subsection{Independent transfer-ionization and capture--shake-off}  
 
Let us now very briefly consider two uncorrelated mechanisms: 
the independent transfer-ionization and capture--shake-off.   

According to the first of them transfer-ionization proceeds 
in two independent steps: one electron is captured (transfer) 
and the other one is emitted (ionization). 
These transitions are driven by the interaction between 
the projectile and the electrons while the electrons 
do not need at all to interact with each other for 
the transitions to occur. Note that within this mechanism 
the projectile must interact with the target at least twice 
(at least one interaction per electron). 

In the consideration of the present paper 
the capture and ionization parts of the independent transfer-ionization 
are regarded as occurring in the collision between a projectile-nucleus 
and a hydrogen-like system. The latter is described using 
an effective nuclear charge which was taken to be $1.69$, 
both for capture and ionization. 
In the impact parameter space the amplitude for this process 
is a product of the single-electron transition 
amplitudes for capture and ionization. The latter ones  
are obtained using the three-body CDW (capture) 
and CDW-EIS (ionization) \cite{cdw-eis} models. 

In capture--shake-off the "instant" removal of one of the electrons 
from the atom due to its capture by the fast projectile  
forces the other electron to react to a sudden 
change of the atomic potential. 
As a result, the second electron can be shaked off 
from the target and become unbound \cite{mcg}.  
The amplitude for this channel is estimated as 
the product of the amplitude for single electron 
capture (evaluated within the three-body CDW -- 
like in case of the ITI) 
and the amplitude for shake-off which is simply an overlap 
between the initial and final states of the ``second'' electron. 

\subsection{The total contribution to transfer-ionization} 

In our calculations we add  
the contributions of the correlated, the independent and 
capture--shake-off channels incoherently. 
In the context of the present paper, 
which is focused on the correlated capture mechanisms, 
such an incoherent addition does not represent 
a big shortcoming since at the collision parameters considered here 
the correlated and uncorrelated have a small overlap 
in the momentum space of the emitted electrons.  

To conclude this section note that the validity 
of our approach to transfer-ionization in fast collisions 
has been already tested in \cite{we-EE}-\cite{ich-EE}   
where the cross sections singly differential in 
the longitudinal momentum of the emitted electrons 
and target recoil ions were calculated for proton 
on helium collisions 
at $v = 12.6$ and $15.2$ a.u. and 
a good agreement with available experimental 
data has been found \cite{foot-new}. 
 
\section{Results and discussion} 

In this section we discuss the momentum spectra for electrons  
emitted in transfer-ionization in collisions of protons, 
alpha-particles and bare lithium nuclei with helium. 

As was mentioned in the previous section, 
in our evaluation of the ITI and C-SO mechanisms 
we use the effective charge of $1.69$ to describe the initial 
undistorted state of the electrons in helium. 
Therefore, for consistency,  in our calculation 
of the contributions from the correlated mechanisms 
we use the set ii) of the parameters 
for the state (\ref{e3}) 
(except in figure \ref{corr-in-init-state}, 
where the sets i) and v) are used). 

\begin{figure}[t] 
\vspace{-0.45cm}
\begin{center}
\includegraphics[width=0.61\textwidth]{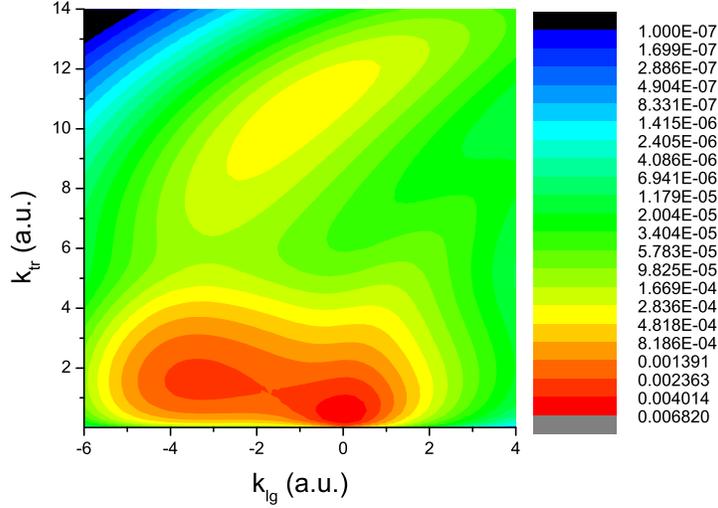}
\end{center}
\vspace{-0.9cm} 
\caption{ \footnotesize{ Momentum spectrum (in b/(a.u.)$^2$) 
of electrons emitted in the reaction    
$3.6$ MeV p$^+$ + He(1s$^2$) $\to$ H(1s) + He$^{2+}$ + e$^-$ 
collisions ($v=12$ a.u.). }}  
\label{12-1} 
\end{figure}

\begin{figure}[t] 
\vspace{-0.45cm}
\begin{center}
\includegraphics[width=0.61\textwidth]{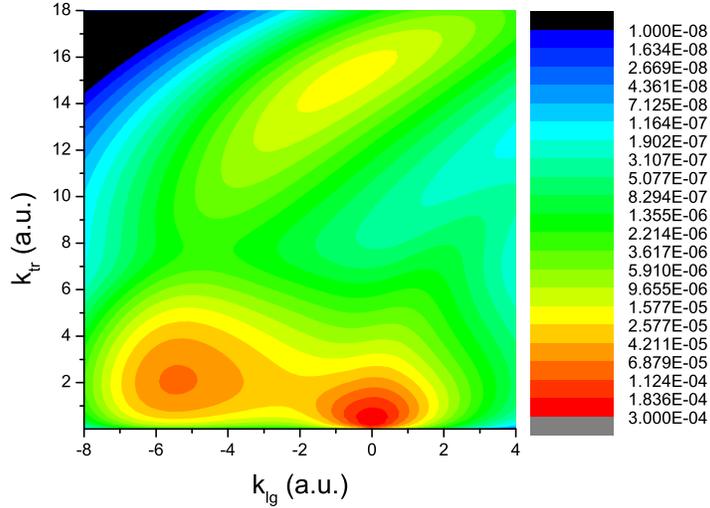}
\end{center}
\vspace{-0.9cm} 
\caption{ \footnotesize{ Same as in figure \ref{12-1} but for 
$6.4$ MeV p$^+$ + He(1s$^2$) $\to$ H(1s) + He$^{2+}$ + e$^-$ 
collisions ($v=16$ a.u.). }}  
\label{16-1} 
\end{figure}

\begin{figure}[t] 
\vspace{-0.45cm}
\begin{center}
\includegraphics[width=0.61\textwidth]{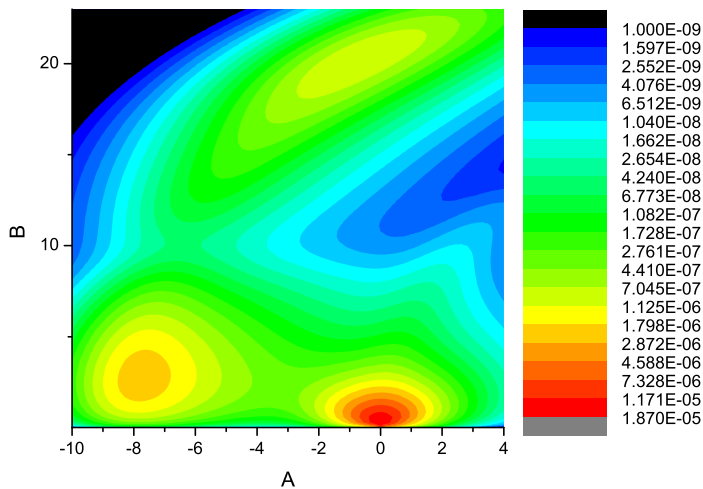}
\end{center}
\vspace{-0.9cm} 
\caption{ \footnotesize{ Same as in figure \ref{12-1} but for 
$11$ MeV p$^+$ + He(1s$^2$) $\to$ H(1s) + He$^{2+}$ + e$^-$ 
collisions ($v=21$ a.u.). }}  
\label{21-1} 
\end{figure}

The momentum spectra shown in figures 
\ref{12-1}-\ref{21-3} are given 
in the rest frame of the target 
(laboratory frame) and are represented 
by the doubly differential cross section 
\begin{eqnarray} 
\frac {d^2 \sigma }{ dk_{lg} dk_{tr} } =  
k_{tr} \int_0^{2 \pi} d\varphi_k \,  
\int d^2 {\bf q}_{\perp} \left| S_{fi}({\bf q}_{\perp}) \right|^2,        
\label{e11} 
\end{eqnarray}  
where $k_{lg} = {\bf k} \cdot {\bf v} /v $ and 
${\bf k}_{tr} = {\bf k} - k_{lg} {\bf v} /v  $ 
are the longitudinal and transverse parts, 
respectively, of the momentum ${\bf k}$  
of the emitted electron. The integration 
in (\ref{e11}) runs over the transverse part of 
the momentum transfer and the azimuthal angle 
$\varphi_k$ of the emitted electron. 
In the range of collision parameters considered here 
an atomic electron is mainly captured 
into the ground state of the projectile. 
Therefore, in what follows we consider  
transfer-ionization only for this channel. 

The momentum spectra of electrons emitted in collisions 
with protons are displayed in figures \ref{12-1}, \ref{16-1} 
and \ref{21-1} for impact energies of $3.6$, $6.4$ and $11$ MeV, 
respectively. These energies correspond to $v=12$, $16$ and $21$ a.u.
It is seen in the figures that there are three distinct 
maxima in the spectra.   

\subsection*{Uncorrelated transfer-ionization} 

The maximum, which is located at small values of $k$, 
has its origin in the uncorrelated mechanisms: 
the independent transfer-ionization and 
capture--shake-off. 

In high-velocity collisions ($v \gg Z_p, Z_t$) 
the cross section for single electron capture 
calculated within the CDW approximation scales 
approximately as $Z_p^5/v^{11}$. In our model, this is obviously 
also the scaling for the contribution  
of the capture--shake-off channel to the cross section.  

Since the ionization part of the independent transfer-ionization 
adds the factor $Z_p^2/v^2$, the cross section for this channel 
is proportional to $Z_p^7/v^{13}$ and, compared to 
the capture--shake-off, shows 
a steeper dependence both on the projectile 
charge and collision velocity. 
 
According to our model, 
in collisions with protons 
(in the range of impact velocities 
considered) the maximum at small $k$ is 
dominated by capture--shake-off, that leads 
to the shape of the spectrum almost symmetric 
with respect to $k_{lg} = 0$ \cite{asym-shake-off}. 
The situation becomes somewhat different in collisions with 
alpha-particles and lithium nuclei 
in which the independent transfer-ionization 
becomes relatively 
more important and, as a result, 
the emission spectrum acquires 
a slight forward-backward asymmetry 
with more emitted electrons 
moving in the forward semi-sphere 
(see figures \ref{16-2}, \ref{21-2} and \ref{21-3}). 

\subsection*{Correlated transfer-ionization} 

The maximum at large (negative) $k_{lg} $ 
appears due to the EEA mechanism whereas the maximum 
at large $ k_{tr} $ is a signature of the EET channel. 

\vspace{0.25cm}  

i) Let us consider the kinematics of these two correlated 
channels of transfer-ionization. To this end it is convenient  
to go first to the rest frame of the projectile-nucleus.  
In this frame the latter particle does not take part 
in the energy balance of the process (because it is heavy 
and is initially at rest). 
Therefore, the energy balance can be written as 
$u_e^2/2 + \Delta E \approx v^2$.  
Here, $u_e$ is the velocity of the emitted electron, 
$\Delta E  = v \Delta Q_{lg}$ is the change in 
energy of the nucleus of the atom with 
$\Delta Q_{lg}$ being the change in 
its longitudinal momentum, $v^2$ is the initial energy 
of the two incident electrons  
and we have neglected the initial and final binding 
energies since $v \gg Z_p$ and $v \gg Z_t$. 
Thus, the velocity $u_e$ of the emitted electron 
in the projectile frame is approximately given by 
\begin{eqnarray} 
u_e = v \sqrt{ 2 (1 - \Delta Q_{lg}/v) }. 
\label{velocity} 
\end{eqnarray}  

\begin{figure}[t] 
\vspace{-0.45cm}
\begin{center}
\includegraphics[width=0.61\textwidth]{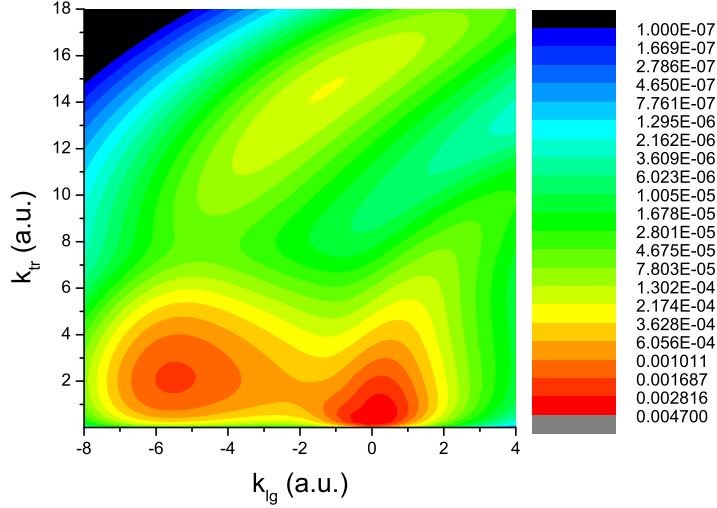} 
\end{center}
\vspace{-0.9cm} 
\caption{ \footnotesize{ Same as in figure \ref{12-1} but for 
$6.4$ MeV/u He$^{2+}$ + He(1s$^2$) $\to$ He$^{+}$(1s) + He$^{2+}$ + e$^-$ 
collisions ($v=16$ a.u.). }}  
\label{16-2} 
\end{figure} 

\vspace{0.25cm} 

ii) If the nucleus of the atom would be just a spectator 
in the collision (and thus $ \Delta Q_{lg} = 0 $), 
one would obtain $u_e \approx \sqrt{2} v$. 
Taking into account that in the target frame the electron emitted via 
the EEA mechanism moves in the direction, which is opposite 
to the projectile velocity, the momentum spectrum of electrons 
produced via the EEA should then be centered in this frame around 
$k_{lg} \approx v - \sqrt{2} v \approx - 0.4 v $. 
Looking at the figures one sees, however, 
that only at the highest impact energy considered ($v=21$ a.u.) 
the electron spectrum is really having the maximum 
at the longitudinal momentum rather close to $- 0.4 v$ while 
at the lower velocities ($v=12$ and $16$ a.u.) 
this maximum is located at a noticeable distance 
from the point $k_{lg} = - 0.4 v$. 
This means that only at sufficiently 
high impact energies the EEA becomes (almost) 
purely electronic mechanism without 
the involvement of the nucleus of the atom.   
At lower impact velocities the target nucleus 
does noticeably participate in this mechanism (see also \cite{ich-EE}).  
  
\begin{figure}[t] 
\vspace{-0.45cm}
\begin{center}
\includegraphics[width=0.61\textwidth]{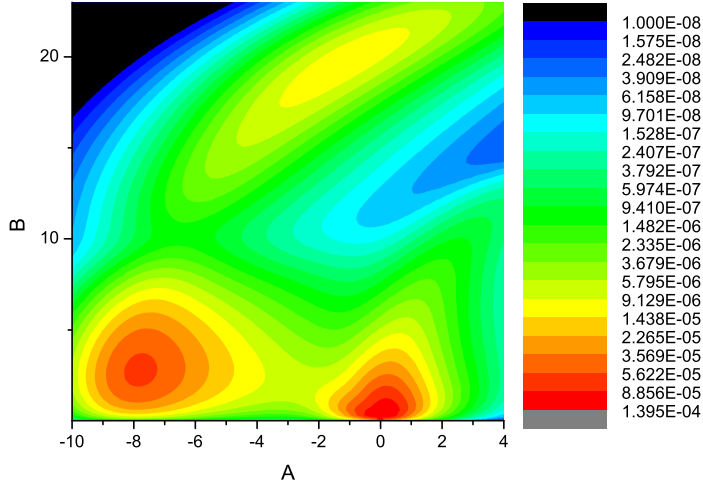}
\end{center}
\vspace{-0.9cm} 
\caption{ \footnotesize{ Same as in figure \ref{12-1} but for 
$11$ MeV/u He$^{2+}$ + He(1s$^2$) $\to$ He$^{+}$(1s) + He$^{2+}$ + e$^-$ 
collisions ($v=21$ a.u.). }}  
\label{21-2} 
\end{figure} 

\begin{figure}[t] 
\vspace{-0.45cm}
\begin{center}
\includegraphics[width=0.61\textwidth]{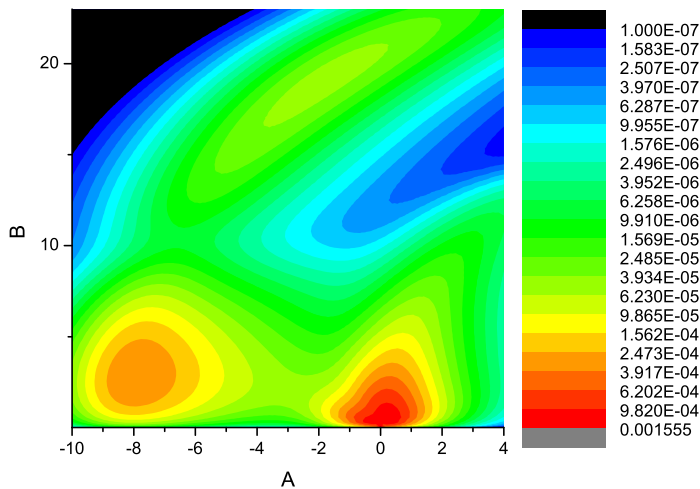}
\end{center}i)
\vspace{-0.9cm} 
\caption{ \footnotesize{ Same as in figure \ref{12-1} but for 
$11$ MeV/u Li$^{3+}$ + He(1s$^2$) $\to$ Li$^{2+}$(1s) + He$^{2+}$ + e$^-$ 
collisions ($v=21$ a.u.). }}  
\label{21-3} 
\end{figure} 

\vspace{0.25cm}  

iii) Following the simple picture of the EET mechanism, 
which was mentioned in the Introduction, one would expect that 
in the rest frame of the target atom the EET is 
characterized by electrons emitted with a velocity $v$ under 
the angle of $90^0$ with respect to the motion of the projectile. 
Correspondingly, the velocity of the electron with respect 
to the projectile should be equal to $\sqrt{2} v$. 
The latter value indeed 
agrees with the simple estimate for the electron velocity 
$u_e$ given above in this subsection 
(if we assume that $ \Delta Q_{lg} = 0 $). 

However, according to the spectra shown in figures 
\ref{12-1}-\ref{21-3}, in the target frame 
the velocity $v_{EET}$ of the emitted electron is on average 
slightly smaller than $v$. Besides, 
the angle $\vartheta_{EET}$, which characterises 
the position of the ``center of mass'' of the EET maximum 
in the momentum spectrum, is somewhat larger than $90^0$: 
$\vartheta_{EET} = 90^0 + \delta$, 
where $\delta > 0$. Moreover, the differences 
$(v - v_{EET})$ and $\delta $ increase  
with increasing the charge of the projectile 
and/or decreasing the impact velocity.   
The reason for this is that the simple picture does not 
take into account that the electron which moves together 
with the projectile is actually not free but bound \cite{foot_note_2}. 
When the binding of the captured electron increases  
the differences between the result and 
what the simple picture suggests also growth. 
Only in the limit $v \to \infty$ does the position of 
the EET maximum coincide with the prediction 
of this picture (see also \cite{briggs}). 

Assuming that at sufficiently high impact velocities 
the nucleus of the target atom is a spectator in the EET mechanism,  
one can find a simple relation between 
the averaged velocity $v_{EET}$ of the emitted electron 
and the angle $\delta $. 
Indeed, in the rest frame of the projectile the energy 
of this electron is approximately given by 
$v^2$. Taking into account that the same energy 
can also be expressed as 
$\left( v + v_{EET} \sin \delta \right)^2/2 + (v_{EET} \cos \delta)^2/2 %
= v^2/2 + v_{EET}^2/2 + v v_{EET} \sin \delta$,  
one obtains $v_{EET} \approx %
v \left(\sqrt{1+\sin^2 \delta} - \sin \delta \right) %
 \approx v \left(1 - \delta \right)$.  

\vspace{0.25cm}  

iv) Two more observations can be drawn from figures \ref{12-1}-\ref{21-3}. 
First, for a fixed projectile charge state 
the relative importance of the EET versus EEA increases with $v$. 
Second, for a fixed collision velocity $v$ 
the EEA mechanism gains in relative importance  
when the charge of the projectile increases. 

The first observation can be understood noting that 
the EEA and EET are basically the first 
and second order processes, respectively \cite{ich-EE}, \cite{briggs}. 
As a result, the EET weakens more slowly with increasing $v$
than the EEA. Further, the dependence of 
the EEA mechanism on $Z_p$ is a bit steeper ($ \sim Z_p^5$) \cite{ich-EE} 
than that of the EET ($ \sim Z_p^5/(Z_p + Z_t \sqrt{2}) $) \cite{briggs}. 
This enables one to understand the second observation. 


\subsection*{ Dynamic versus stationary correlations } 

Both the EEA and EET mechanisms crucially rely on 
the coulomb interaction between the electrons. 
On the other hand, the electron-electron correlations  
in the initial and final asymptotic states 
of the colliding system are also manifestation 
of this interaction \cite{corr-in-init-state}. 

The principal difference between them 
is that while the correlations in 
the asymptotic states are stationary in nature, 
the EEA and EET are based on 
the electron-electron interaction 
in its dynamic variant. 

An illustrative example of the correspondence between 
dynamic and stationary manifestations of basically the same force  
is represented by the interaction between an electron 
and its electromagnetic field (the radiation field).  
A stationary situation is realized, 
for instance, when one considers  
a free (undistorted) hydrogen atom in the ground state. 
In this case the interaction with the radiation field 
has quite a weak impact on the system: it merely leads 
to a tiny shift of the energy of the ground state. 
Let us consider, however, a situation when the hydrogen atom 
collides with a fast ion. Now the same interaction 
may lead to electron transfer from the atom to 
the ion, which is called radiative electron capture. 
In this dynamic situation the interaction with 
the radiation field leads to a drastic change  
in the state of the electron.  

The difference between the stationary and 
dynamic manifestations of the electron-electron interaction 
is not that dramatic. Nevetheless, it is the dynamic electron 
correlations (the EEA and EET),  
which drive the process of transfer-ionization,   
whereas the stationary ones providing an ``environment'' 
also influence the process by determining, for instance,  
the mean electron-electron distance 
in the initial atomic state 
and, thus, the magnitude of the dynamic force 
acting between the electrons in 
the transfer process \cite{corr-final-state}. 

\begin{figure}[t] 
\vspace{-0.45cm}
\begin{center}
\includegraphics[width=0.71\textwidth]{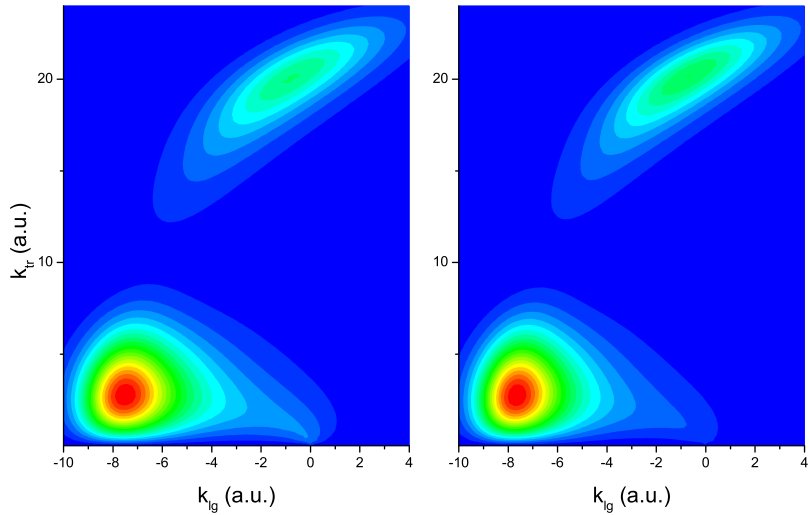}
\end{center}
\vspace{-0.9cm} 
\caption{ \footnotesize{ The calculated contribution 
of the EEA and EET mechanisms to the momentum spectrum 
of electrons emitted 
in $11$ MeV p$^+$ + He(1s$^2$) $\to$ H(1s) + He$^{2+}$ + e$^-$ 
collisions ($v=21$ a.u.). The left panel: 
$\alpha = \beta = 2$, $\gamma = 0$. 
The right panel: $\alpha=2.21$, $\beta=1.44$ 
and $\gamma=0.207$. }}  
\label{corr-in-init-state} 
\end{figure}

These points can be seen 
in figure \ref{corr-in-init-state} 
where we present the contributions 
to the momentum spectrum  
due to the EEA and EET mechanisms 
calculated with two different approximations 
for the ground state of helium.   
In the left plot the parameters 
of the state (\ref{e3} ) were taken as  
$\alpha = \beta = 2$ and $\gamma = 0$ which means 
that the electron-electron correlations 
in this state are completely ignored.  
The right plot was obtained by choosing 
$\alpha=2.21$, $\beta=1.44$ and $\gamma=0.207$ 
which includes (in an approximate way) 
both the radial and angular correlations 
in the ground state of helium. 
It is seen that there is practically no difference 
in shape of these two spectra. However, 
their absolute intensities 
differ by about a factor of $2$ 
since the electron-electron interaction  
in the ground state of helium increases 
the mean electron-nucleus distances in the atom 
and, thus, decreases the effective strength 
of the EEA and EET channels (see also \cite{ich-EE}). 

\subsection*{Kinematics of the uncorrelated transfer-ionization channels} 

To conclude our discussion in this section note 
that Eq. (\ref{velocity}) is of course also 
valid for the uncorrelated mechanisms. 
In contrast to the correlated ones, however, 
in this case we have $u_e \approx v$ 
and, hence, $ \Delta Q_{lg} \approx v/2 $. 
Therefore, it is the nucleus of the atom which 
balances (in the projectile frame) 
the energy change of the captured electron,  
$\Delta E  = v \Delta Q_{lg} \approx v^2/2$, 
both in the independent transfer-ionization and 
capture--shake-off channels. 

\section{Conclusions} 

We have considered in some detail transfer-ionization 
in collisions of fast protons, alpha-particles 
and lithium nuclei with atomic helium. 
There are four basic mechanisms which are responsible 
for this process. Two of them (the independent transfer-ionization 
and capture--shake-off) are so called uncorrelated mechanisms 
which means that they would not disappear if 
the electron-electron interaction would be ``switched off''.   
In contrast, this interaction does play a crucial role 
in the other two (the electron-electron and electron-electron-Thomas)  
mechanisms which both are governed by the  
dynamic electron-electron correlations. 

Our consideration shows that at sufficiently 
high impact velocities the contributions of 
the correlated and uncorrelated mechanisms 
can be clearly separated in the cross section doubly 
differential in the longitudinal and transverse 
components of the momentum of the emitted electron. 
The study of this cross section also enables one 
to separate the two correlated mechanisms 
from each other and get insight into 
subtle details of the dynamics of transfer-ionization.  

At high impact energies $v \gg Z_t, Z_p$ 
the position of the center of the maximum in the momentum spectrum, 
caused by the EEA mechanism,  
tends in the target frame to $k_{lg} = -0.4 v$.  
This means that the role of 
the nucleus of the atom in this mechanism weakens with increasing 
collision velocity and the EEA eventually becomes 
a truly electronic one. However, according to our model, 
even at impact velocities as high as $12$ and $16$ a.u. 
the helium nucleus still noticeably participates in this process. 

According to the well known picture of the EET mechanism 
the emitted electron should have a velocity equal 
to the collision velocity $v$ and fly under 
the angle $90^0$ with respect to the projectile motion. 
Our model predicts, however, that the velocity 
of the emitted electron is on average smaller than $v$ and 
that the electron is emitted under the angle which 
is larger than $90^0$. These two differences are interconnected 
and increase if the charge of the projectile increases 
and/or the impact velocity decreases. 

An experimental exploration of the spectra of electrons 
emitted in transfer-ionization at high impact 
velocities is very desirable. 
At the highest velocity ($v = 21$ a.u.), considered in this article, 
the total cross section for transfer-ionization, 
according to our estimates,  
is of the order of $0.1$, $1$ and $10$ mb in collisions with protons, 
alpha-particles and lithium nuclei. These values are of course 
rather small. Note, however, that already several years ago 
it was possible (see \cite{schmidt}) to measure  
the longitudinal momentum spectrum of the recoil target ions 
for transfer-ionization process with the total cross section 
of the order of $1$ mb. 

\section*{Acknowledgement} 

A.B.V. acknowledges the support from the Extreme Matter Institute EMMI 
and the program for visiting international senior scientists 
of Chinese Academy of Sciences.


\begin{thebibliography}{99}
\bibitem{foot_note} In general, using the rest 
frame of the ion this process  
can be viewed as a kind of "Auger decay" 
in which one of the electrons  
initially belonging to a fast moving configuration 
of atomic particles undergoes a radiationless transfer 
into a bound state of the ion.
\bibitem{foot-new} Moreover,  
even at relatively low impact velocities 
($Z_p/v \simeq 0.5$)  
this approach can still yield reasonable results 
(see \cite{daniel}).
\bibitem{asym-shake-off} When the angular correlations 
are included in the ground state of helium, 
the calculated contribution of the C-SO channel 
to the emission spectrum becomes slighly asymmetric: 
the maximum of the spectrum is located at 
$p_{lg} = 0 $ but more electrons are 
emitted in the direction opposite to the projectile motion.
\bibitem{foot_note_2} Neither this picture 
takes into account that the electrons 
are initially bound in the atom.
\bibitem{corr-in-init-state} 
A possible effect of electron correlations 
in the ground state of helium on transfer-ionization 
in fast proton-helium collisions had recently attracted 
quite a bit of attention \cite{mergel}-\cite{schmidt}.
\bibitem{corr-final-state} In our case 
the electron-electron correlations in 
the final state are weak due to a very high relative velocity 
($ \simeq \sqrt{2} v$) between the electrons.
\end{thebibliography}
\end{document}